\documentclass[]{interact}
\usepackage{epstopdf}
\usepackage{subfigure}
\usepackage[numbers,sort&compress,merge]{natbib}
\usepackage{graphicx}
\usepackage{amsmath,amssymb}
\usepackage{braket}
\usepackage{array}
\usepackage{subfig}
\usepackage{makecell}

\bibpunct[, ]{(}{)}{,}{n}{,}{,}

\renewcommand\bibfont{\fontsize{10}{12}\selectfont}
\renewcommand\citenumfont[1]{\textit{#1}}
\renewcommand\bibnumfmt[1]{(#1)}

\theoremstyle{plain}

\theoremstyle{definition}

\theoremstyle{remark}

\begin{document}

\articletype{}

\title{Single photon sources with different spatial modes}

\author{
\name{Nijil Lal\textsuperscript{a,b}\thanks{CONTACT Nijil Lal. Email: nijil@prl.res.in.} Anindya Banerji\textsuperscript{a}, Ayan Biswas\textsuperscript{a,b}, Ali Anwar\textsuperscript{a,$\dagger$} and R. P. Singh\textsuperscript{a}}
\affil{\textsuperscript{a}Physical Research Laboratory, Ahmedabad 380009, India; \textsuperscript{b}Indian Institute of Technology, Gandhinagar 382355, India; \textsuperscript{$\dagger$}Currently at Centre for Quantum Technologies, National University of Singapore, 3 Science Drive 2, S117543, Singapore}
}

\maketitle

\begin{abstract}
We study the correlation properties of single photons carrying orbital angular momentum (OAM) in a Hanbury Brown and Twiss (HBT) type experiment. We have characterized single photon sources obtained by pumping a nonlinear crystal with a laser beam carrying different OAM under same experimental conditions. For heralded twisted single photons carrying OAM, we calculate $g^{(2)}(0)$, a measurable parameter characterizing the quality of a single photon source, and observe an increment with the OAM of the single photon.
\end{abstract}

\begin{keywords}
Single photon sources, Optical vortices, Spatial modes of light, Intensity correlation.
\end{keywords}

\section{Introduction}

Single photon sources are one of the most important quantum sources of light finding applications in quantum key distribution, random number generation, quantum computing with photons, and quantum metrology \cite{loqkd,herrerornd,northup,giovannettimetrology}. One of the most popular technique to produce a single photon source is to use spontaneous parametric down conversion (SPDC) process in a $\chi^{(2)}$ nonlinear crystal \cite{boydbook,agarwalbook}. In this process, one photon of the pump is converted to two photons of lower energies, which propagate in certain direction following the conservation laws of energy and momentum. Since the two down converted photons are generated at the same time, detection of one photon heralds the presence of the other \cite{burnham,hong1986heralding}. Therefore, single photon sources obtained by using this technique are generally called as heralded single photon sources.
Optical vortices or Laguerre Gaussian (LG) beams with zero radial index are gaining popularity in implementing various quantum information protocols \cite{mirhosseini2015,barreiro} as they provide extra degree of freedom in the form of orbital angular momentum (OAM) \cite{allen,yao2011} that can be measured using standard experimental techniques \cite{pravin,shashi,mair}. Optical vortices of different orders or azimuthal indices form different spatial modes of light with their characteristic properties \cite{ashokcorr,ashokinfo, reddyspeckle,reddydivergence,kedar}. Intensity correlations of classical optical vortices are found to have dependence on the order of the vortex, while scattered from a rotating ground-glass plate \cite{ashokcorr,ashokhbt}. Such studies evoke interest in the correlation properties of single photons carrying OAM generated in spontaneous processes such as parametric down conversion. In the present work, starting with Gaussian mode, we take different order of vortices as a pump to the nonlinear crystal and study the intensity correlations of down converted photons to characterize them for single photon sources of light including twisted single photons.

\section{Theory}

\begin{figure}[h]
    	\centering
    	\begin{minipage}[h]{0.4\textwidth}
    		\centering
    		\includegraphics[scale=0.7]{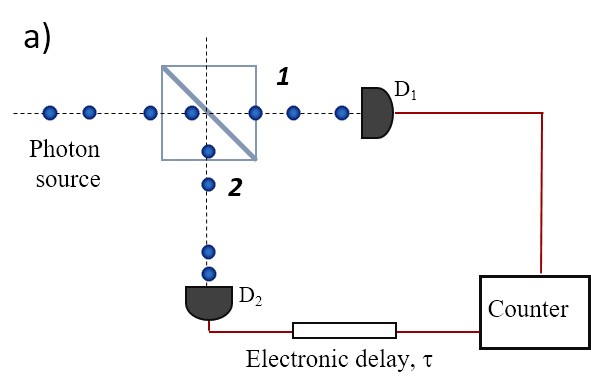}
    	\end{minipage}
    	\hspace*{\fill}
    	\begin{minipage}[h]{0.4\textwidth}
    		\centering
    		\includegraphics[scale=0.7]{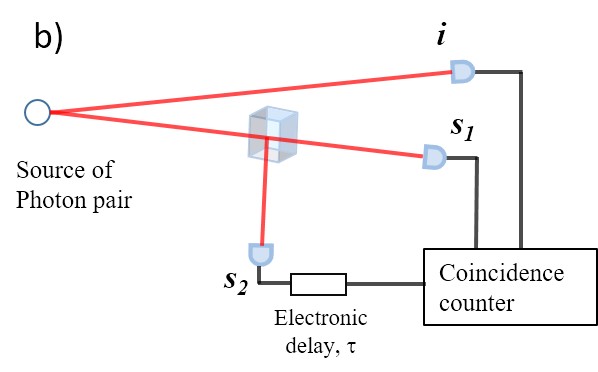}
    	\end{minipage}
    	\hspace*{\fill}
    	\caption{A simplified illustration of HBT experiment for a) an ideal single photon source where photons reaching the beam splitter through the input port will choose any of the paths $1$ or $2$ without resulting in any coincidence detection between the detectors, $D_1$ $\&$ $D_2$ and b) a heralded single photon source.}
    	\label{fig:HBT}
    \end{figure}

The measurement of second order correlation ($g^{(2)}(\tau)$) has been important in quantum optics, especially in observing antibunching which is a purely quantum mechanical behaviour. The normalized second order correlation function \cite{mandelwolf}, for a fixed position, is given as,
\begin{equation}
    g^{(2)}(\tau) =\frac{\braket{ \hat{E}^{(-)}(t)\hat{E}^{(-)}(t+\tau)\hat{E}^{(+)}(t+\tau)\hat{E}^{(+)}(\tau)}}{\braket{ \hat{E}^{(-)}(t)\hat{E}^{(+)}(t)}\braket{ \hat{E}^{(-)}(t+\tau)\hat{E}^{(+)}(t+\tau)}}
    \label{eqn:g2c}
\end{equation}

and is usually measured in a Hanburry Brown-Twiss (HBT) experiment \cite{kimble,u'ren}, which gives the correlation between two photons separated by a time delay, $\tau$. For an ideal single photon source from which individual photons are emitted one after another as given in Figure \ref{fig:HBT}a, a photon either gets transmitted (along arm $\textbf{\textit{1}}$) at the beam splitter or gets reflected (along arm $\textbf{\textit{2}}$). There are no simultaneous incidences at the detectors $D_1$ $\&$ $D_2$ when there is no delay between the paths $\textbf{\textit{1}}$ and $\textbf{\textit{2}}$ ($\tau = 0$). For heralded single photon sources generated by spontaneous parametric down conversion (SPDC), the photon correlation in the HBT experiment has to be done in the signal ($\textbf{\textit{s}}$) with the conditioned detection of idler($\textbf{\textit{i}}$) (Figure \ref{fig:HBT}b). The second order correlation function for this configuration will take the form \cite{grangier,u'ren},

\begin{equation}
    g^{(2)}(0) = \frac{R_{i,s_1,s_2} R_{i}}{R_{i,s_1} R_{i,s_2}}
    \label{eqn:g2_3d}
\end{equation}
where $R_{i}$ is the count rate of singles in the heralding arm (\textbf{\textit{$i$}}), $R_{i,s_1}$ and $R_{i,s_2}$ are the respective two-fold coincidence rates between \textbf{\textit{$i-s_1$}} and \textbf{\textit{$i-s_2$}}, and $R_{i,s_1,s_2}$ is the rate of triple coincidences between \textbf{\textit{$i-s_1-s_2$}}.
    
A modified version of this equation takes the form \cite{beck},

\begin{equation}
g^{(2)}(0)=R_{i} \Delta t(R_{s_1}/R_{i,s_1} +R_{s_2}/R_{i,s_2} )  
\label{eqn:g2_beck}
\end{equation}
with the assumption that simultaneous detection in all three detectors results from accidental coincidence in one of the signal arms along with a true two-fold coincidence the other two. Here, the coincidence window is $ \Delta t$.

\section{Experiment}
\begin{figure}[h]
\begin{center}
\includegraphics[scale=0.6]{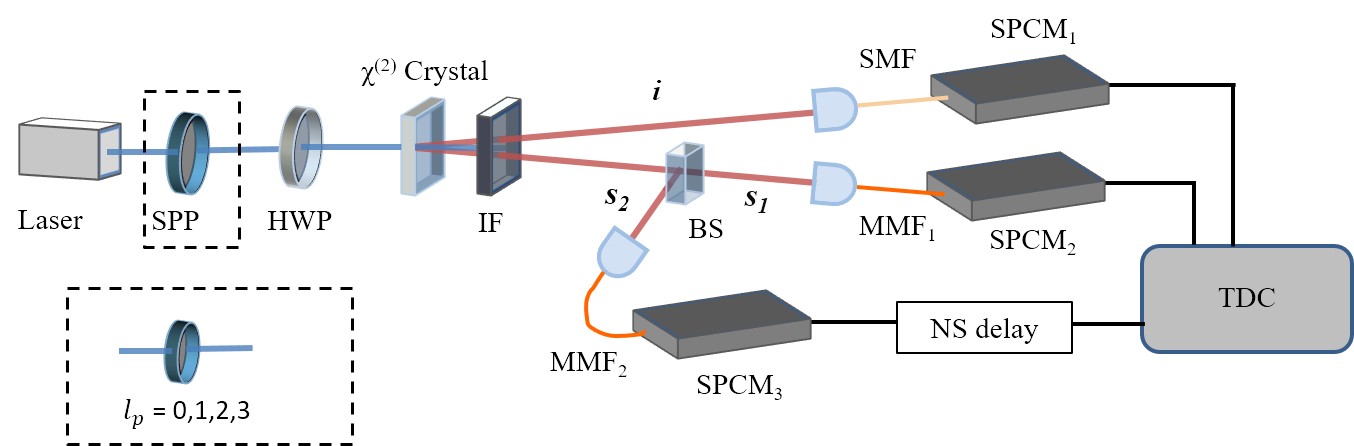}
\caption{HBT-like setup to determine the second order correlation for a heralded single photon source. The delay generator introduces electronic delay in steps of 0.5 ns between the two output ports of the beam splitter (BS) in the signal arm.}
\label{fig:setup}
\end{center}
\end{figure}

Experimental setup for studying the correlation of heralded signal photons from the SPDC is given in Figure \ref{fig:setup}. A diode laser (Toptica Topmode) of wavelength 405 nm is used as pump. Optical vortices of different orders, generated using spiral phase plates (SPP) (Holo/Or), are incident on a $\chi^{(2)}$ non-linear crystal ($\beta$-Barium Borate or BBO, 5x5x2 mm), which generates a cone of correlated pairs of down-converted photons at phase-matching conditions. Degenerate signal-idler pairs are selected using interference filters. The pump polarization is adjusted along crystal optic axis using a half wave plate ensuring maximum down-conversion. The idler photons are coupled to a single-mode fiber due to which only photons with OAM, $l_i = 0$ are selected in the idler arm. This ensures that the corresponding signal photons are carrying same OAM as the pump photon following the OAM conservation. The signal is equally split using a 50:50 beam splitter and each portion is coupled to a multimode fiber. The idler and signals from the two ports of the beam splitter are then guided to single photon counting modules (Excelitas SPCM-AQRH-16-FC, dark counts $\sim$ 25 cps). The interference filters (IF) of pass band 810 $\pm5$ nm cut the undesired light at detection. Optimizing fiber coupling using fiber collimators, maximum coincidences are achieved between signal and idler arms.
The coincident photons are counted using a time to digital converter (ID800 TDC, IDQuantique). The coincidence window for heralding is kept as 410 ps. The detector positions are adjusted such that the relative delay between all three detectors are zero. This is ensured by obtaining maximum two fold coincidences between $\textbf{\textit{$i$}}$ - $\textbf{\textit{$s_1$}}$ and $\textbf{\textit{$i$}}$ - $\textbf{\textit{$s_2$}}$.

\section{Results and Discussion}

It is known that orbital angular momentum is conserved in parametric down conversion \cite{mair} such that the sum of the orbital angular momenta of the down converted photons equals the pump OAM. While the output beams of the parametric down conversion are incoherent sum of multiple spatial modes, the signal-idler photon pairs will have OAM $l_s$ and $l_i$ respectively, restricted by the OAM conservation  $l_i + l_s = l_p$, where $l_p$ is the pump OAM. \textit{Projecting the idler into a single spatial mode (Gaussian, i.e. $l_i = 0$) by coupling it to a single mode fiber, the OAM of the signal is selected to be the same as that of the pump following the OAM conservation}. Hence, the heralded correlation measurements done in the signal arm give the statistics of twisted photons whose OAM is defined by that of the pump. The pump OAM is varied using SPPs of different orders to study the statistics of generated single photons. Heralded second order correlation is measured for each order of pump OAM.

\begin{figure}[h]
\begin{center}
\includegraphics[scale=0.3]{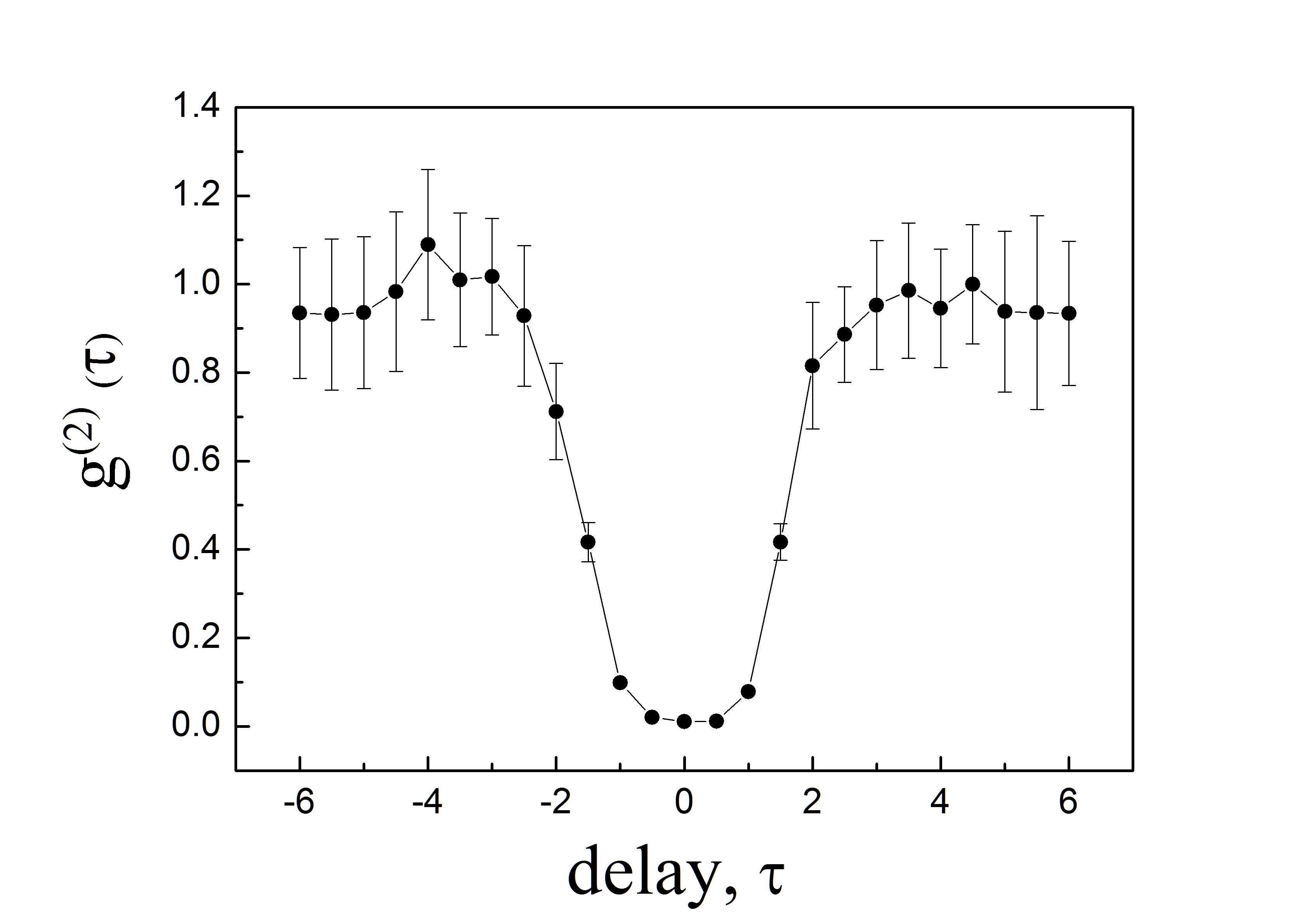}
\caption{The variation of second order correlation function with delay, $\tau$}
\label{fig:g2tau}
\end{center}
\end{figure}

At first, the second order coherence function, $g^{(2)} (\tau)$, is calculated from Equation (\ref{eqn:g2_3d}) as a function of the time delay between photons reaching the beam splitter. A delay generator introduces electronic delay in steps of 0.5 ns between the two output ports of the beam splitter ($\textbf{\textit{$s_1$}}$ $\&$ $\textbf{\textit{$s_2$}}$) in the signal arm. Coincidence measurements recorded for pair of detectors $SPCM_1$ $\&$ $SPCM_2$ kept in these two arms are heralded by the detection of a photon in the idler arm $\textbf{\textit{i}}$. Antibunching (for $\tau = 0$) is observed by varying the temporal delay between arms \textbf{\textit{$s_1$}} and \textbf{\textit{$s_2$}} as given in Figure \ref{fig:g2tau}.

\begin{figure}[h]
\begin{center}
\includegraphics[scale=0.3]{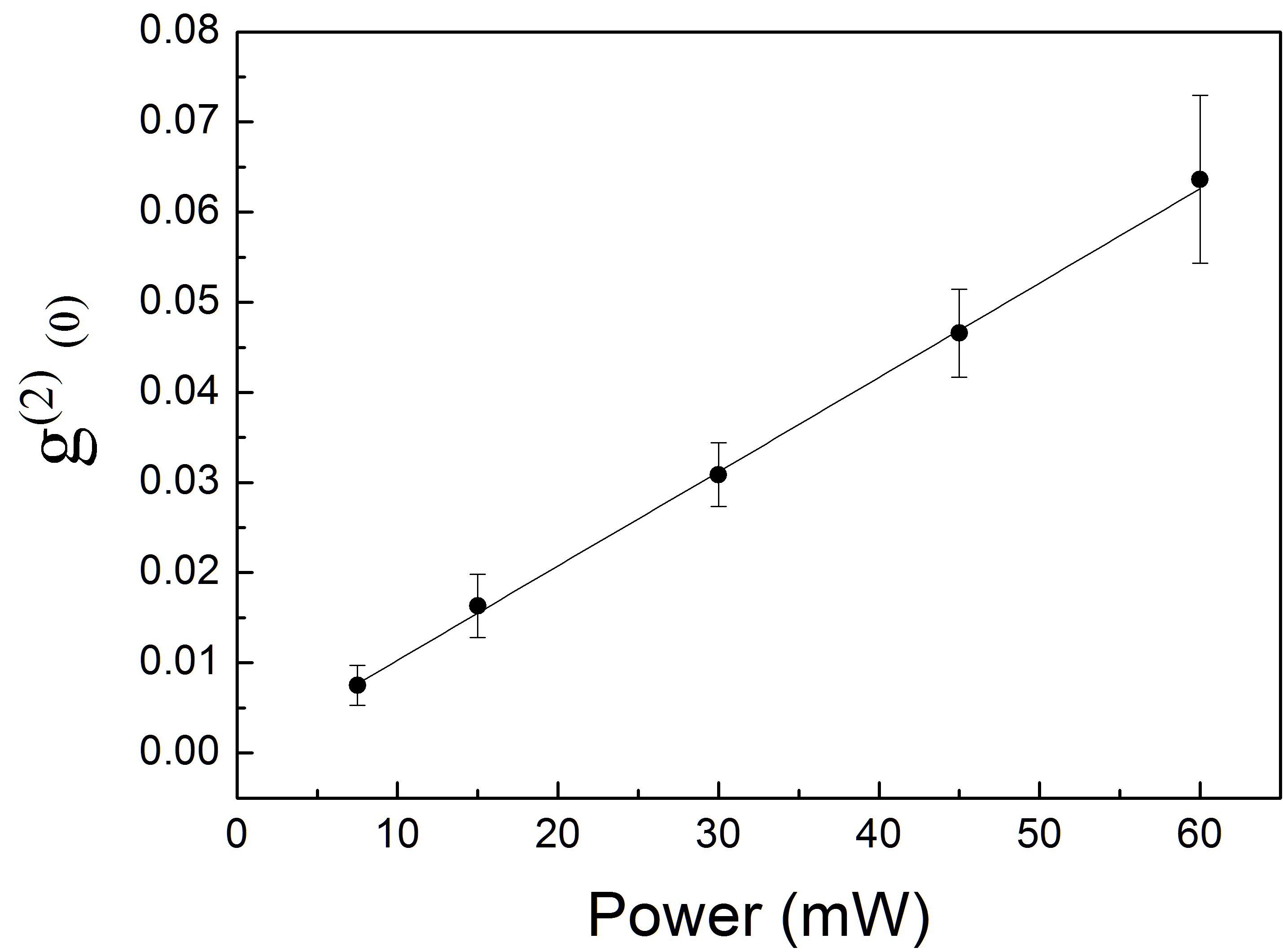}
\caption{The variation of $g^{(2)} (0)$ with pump power}
\label{fig:g2power}
\end{center}
\end{figure}

While the expected value of  $g^{(2)} (0)$ for a true single photon source is zero, the non-zero values in a heralded detection can be attributed to various reasons such as accidental coincidences, simultaneous generation of multiple photon pairs. We have studied $g^{(2)} (0)$ by increasing the pump power and observed an increase with power(Figure \ref{fig:g2power}). The increment in $g^{(2)} (0)$  with pump power can be attributed to the fact that the rate of pair production and hence the probability of simultaneous creation of multiple photon pairs are proportional to the pump power \cite{razavi}. This results in more than two photons reaching together at the beam splitter leading to coincidences between the two signal arms. A higher  $g^{(2)} (0)$ shows a reduced non-classical behaviour.  Hence, it is crucial for quantum optical experiments using heralded single photon sources to be done in the lower pump power regime.

The second order correlation function for $\tau = 0$, $g^{(2)}(0)$, measured in low power regime (pump power = 7 mW) for different OAM values of single photons is given in Table \ref{tab:g20}. For a single photon with $l = 0$, we have obtained $g^{(2)}(0)$ = 0.0082 $\pm$ 0.0043 from Equation (\ref{eqn:g2_3d}). As the probability of detection of three-fold coincidences is very low, the statistical error in such measurements can be relatively high. Hence, we have also calculated $g^{(2)}(0)$ using Equation (\ref{eqn:g2_beck}), where the direct measurements of three-fold coincidences are not used. Here, $g^{(2)}(0)$ is calculated with the assumption that three-fold coincidences result from the simultaneous occurrence of a two-fold coincidence (between the heralding photon and one of the signal photons) and an accidental coincidence in the third detector. In both cases, an increment in $g^{(2)}(0)$ has been observed with the increasing orders of the OAM.

\begin{table}[h]
    \centering
\begin{tabular}{ | m{4cm} | m{2cm}| m{2cm} | m{2cm} | m{2cm} | } 
\hline
\thead{Twisted photon \\ OAM, \textit{l}}& \thead{0} & \thead{1}& \thead{2}& \thead{3} \\ 
\hline
\thead{a) \hspace{0.5cm} $g^{(2)}(0)$}  & \makecell{0.0082\\ $\pm$ 0.0043} & \makecell{0.015\\ $\pm$ 0.011} & \makecell{0.030\\ $\pm$ 0.057} & \makecell{0.045\\ $\pm$ 0.169} \\
\hline
\thead{b) \hspace{0.5cm} $g^{(2)}(0)$  } & \makecell{0.0047\\ $\pm$ 0.0001} & \makecell{0.0094 \\$\pm$ 0.0003} & \makecell{0.021\\ $\pm$ 0.001} & \makecell{0.042\\ $\pm$ 0.002} \\
\hline
\end{tabular}
    \caption{Second order correlation, $g^{(2)}(0)$, for a single photon source a) measured directly from three fold coincidences using Equation (\ref{eqn:g2_3d}) and b) accounting only for accidental coincidences using Equation (\ref{eqn:g2_beck}).}
    \label{tab:g20}
\end{table}

The top row in Table \ref{tab:g20} lists $g^{(2)}(0)$ obtained through the direct measurement of triple coincidences between three detectors using Equation (\ref{eqn:g2_3d}) for different orders of OAM of single photons. It can be seen that the error is larger compared to the bottom row since very few triple coincidences are registered even for longer exposure times. This error corresponds to the statistical standard deviation determined using error propagation formula in Equation (\ref{eqn:g2_3d}). The bottom row gives the same but only attributing the non-zero value of $g^{(2)}(0)$ to possible accidental coincidences between the three detectors. Hence we obtain a lower value for $g^{(2)}(0)$ compared to the direct measurement from three-fold coincidences with a lower standard deviation. It is also observed that $g^{(2)}(0)$ is increasing with increasing order of OAM of twisted photons. The increment in $g^{(2)} (0)$ can be due to the reduced coupling of higher order twisted photons to the fiber as the mode size increases with higher orders of OAM \cite{reddydivergence,curtisstructure}.

\section{Conclusion}

We have calculated the second order coherence function with zero delay, $g^{(2)} (0)$ for different OAM values of heralded single photons. It was calculated from the direct measurement of the simultaneous detection between the three detectors in a heralded HBT experiment as well as accounting for the accidental three-fold coincidences. In both cases, $g^{(2)} (0)$ is found to be increasing with the increasing order of the OAM for the twisted photons. The results can be significant while using OAM states of single photons for quantum information applications.

\bibliographystyle{tfp}

\end{document}